# Unusually strong electronic correlation and field-induced ordered phase in YbCo$_2$

**J Valenta**[1*], **N Tsujii**[1*], **H Yamaoka**[2], **F Honda**[3,4], **Y Hirose**[5] **H Sakurai**[6], **N Terada**[7], **T Naka**[8], **T Nakane**[8], **T Koizumi**[3,9], **H Ishii**[10], **N Hiraoka**[10], and **T Mori**[1]

[1] International Center for Materials Nanoarchitectonics (WPI-MANA), National Institute for Materials Science, 1-2-1, Sengen, Tsukuba, Ibaraki 305-0047, Japan
[2] RIKEN Spring-8 Center, Sayo, Hyogo 679-5148, Japan
[3] Institute for Materials Research, Tohoku University, Ōarai, Ibaraki 311-1313 Japan
[4] Central Institute of Radioisotope Science and Safety, Kyushu University, Fukuoka 819-0395, Japan
[5] Department of Physics, Niigata University, Niigata 950-2181, Japan
[6] Center for Green Research on Energy and Environmental Materials, National Institute for Materials Science, 1-1 Namiki, Tsukuba, Ibaraki 305-0047, Japan
[7] Research Center for Advanced Measurement and Characterization, National Institute for Materials Science, Sengen 1-2-1, Tsukuba 305-0047, Japan
[8] Research Center for Functional Materials, National Institute for Materials Science, Sengen 1-2-1, Tsukuba 305-0047, Japan
[9] Graduate School of Engineering, Tohoku University, Sendai 980-8579, Japan
[10] National Synchrotron Radiation Research Center, Hsinchu 30076, Taiwan
E-mail: Jaroslav.VALENTA@nims.go.jp, tsujii.naohito@nims.go.jp

**Abstract.** We report the first study of electrical resistivity, magnetization, and specific heat on YbCo$_2$. The measurements on a single-phased sample of YbCo$_2$ bring no evidence of magnetic ordering down to 0.3 K in a zero magnetic field. The manifestations of low Kondo temperature are observed. The specific heat value divided by temperature, *C/T*, keeps increasing logarithmically beyond 7 J/mol·K$^2$ with decreasing temperature down to 0.3 K without no sign of magnetic ordering, suggesting a very large electronic specific heat. Analysis of the magnetic specific heat indicates that the large portion of the low-temperature specific heat is not explained simply by the low Kondo temperature but is due to the strong intersite magnetic correlation in both the 3*d* and 4*f* electrons. Temperature-dependent measurements under static magnetic fields up to 7 T are carried out, which show the evolution of field-induced transition above 2 T. The transition temperature increases with increasing field, pointing to a ferromagnetic character. The extrapolation of the transition temperature to zero field suggests that YbCo$_2$ is in the very proximity of the quantum critical point. These results indicate that in the unique case of YbCo$_2$, the itinerant electron magnetism of Co 3*d*-electrons and the Kondo effect within the vicinity of quantum criticality of Yb 4*f*-local moments can both play a role.

## 1. Introduction

The interest of Ce, Eu or Yb elements in intermetallic compounds is based on valence instability and *c-f* electron interactions which bring many interesting behaviors such as heavy fermion, non-BCS superconductivity, non-Fermi liquid, valence transition, quantum criticality, etc... These phenomena are understood by considering the strength of *c-f* interaction which influences effectivity of RKKY interaction to promote magnetic ordering, whereas otherwise, the Kondo effect overwhelms RKKY interaction to form a nonmagnetic Fermi-liquid state. The *c-f* hybridization also results in the fact that 4*f*-electrons of these elements can create local or itinerant magnetic moments and in many cases are in a mixed-valence state. These characteristics sometimes favor forming heavy fermion states where the ground states can be easily altered from nonmagnetic Fermi liquid to magnetic states or vice versa by slight perturbations like chemical substitution, pressure, and magnetic fields for example in $CeRu_2Si_2$, $YbCo_2Zn_{20}$, $YbRh_2Si_2$, and some rare-earth borides [1-4]. These properties can be mapped by Doniach's phase diagram. Here, it is assumed that the conduction electrons are free-electron-like with no magnetic correlation, and the coupling between conduction and *f*-electrons, $J_{cf}$, is the single energy parameter that is responsible for the ground state of the materials. However, several compounds possess properties that are difficult to be explained by the above-mentioned scenario. For example, Fermi liquid state is expected to evolve below Kondo temperature ($T_K$), but it is suggested that a different correlation develops well below $T_K$ for several compounds such as $CePd_3$ [5,6], $YbAl_3$ [7], and $Yb_4TGe_8$ [8,9]. In these cases, a simple free-electron picture for the conduction band should not be adequate. Instead, a unique carrier density and/or the correlation of the conduction band especially due to *d*-electrons are also suggested to be important for the complex physical properties in addition to the conventional $J_{cf}$.

Examples of such additional correlation due to *d*-electrons can be found among the $RCo_2$ compounds (*R* represents rare-earth element) which are known for 3*d*-4*f* magnetism. Compounds with Ce, Eu or Yb elements can be involved in the above-mentioned problem. In the literature, detailed research studies can be found for $CeCo_2$ [10-14]. The $CeCo_2$ reflects different properties than other compounds in the $RCo_2$ series, which are strong ferro/ferri-magnets with two magnetic sublattices created by itinerant Co and localized *R* magnetic moments [15-19]. For $CeCo_2$, on the contrary, the Co 3*d* bands strongly hybridized with 4*f*-bands, which drives the compound almost to the nonmagnetic regime both for 4*f* and 3*d* electrons [13]. This brings our attention to the compounds with Eu and Yb in the $RCo_2$ series of which magnetism of 4*f* electrons should be also affected by the 3*d*-electron correlation. In the literature can be found only two papers about $YbCo_2$ [20,21]. They refer to the crystal structure and X-ray absorption measurement.

In this paper, we bring results of characteristic measurements and analysis on successfully-prepared single-phase $YbCo_2$ polycrystalline samples. Long-range magnetic ordering which was previously theoretically predicted at around 5 K [22] is not observed. Instead, the compound remains to be paramagnetic down to at least 0.3 K, and a very large electronic specific heat coefficient above 7 J/mol·K$^2$ is suggested. Furthermore, we observed an indication of a field-induced ordered phase in magnetization, electrical resistivity, and specific heat of $YbCo_2$. This indicates that in $YbCo_2$, the Kondo effect and the intersite magnetic correlation are competing, which places the system in the very vicinity of the quantum critical point.

## 2. Experimental

The YbCo$_2$ was prepared from mixed powders of pure elements Co (3N) and Yb (3N) in stoichiometry 2:1 with an addition of 5% of Yb weight due to evaporation. Pellets of mixed powders were sealed in Ta-tube and quickly heated in an induction furnace under Ar flow up to 1200 ºC for 5 minutes. Subsequently, the Ta-tube with the sample was annealed at 800 ºC for 5 days. The brittle product of YbCo$_2$ was consolidated by using a Spark Plasma Sintering device (SPS, LABOX$^{TM}$-110, Sinter Land). The sample was pressed up to 50 MPa and quickly heated at 900 ºC with keeping the temperature for 5 minutes under a vacuum of ~10$^{-2}$ Pa. The x-ray diffraction (performed on Miniflex, Rigaku equipped with a Cr-target) showed a single phase of YbCo$_2$ and confirmed crystallization in the cubic MgCu$_2$-type Laves phase with lattice parameter $a = 7.1135(2)$ Å (see Figure 1). The stoichiometric composition was confirmed by scanning electron microscope (JSM-7001F, JEOL Ltd.) equipped with energy dispersive spectroscopy (EDS) in results 66.78 atm% of Co and 33.22 atm% of Yb. The sample contained a negligible amount of Yb$_2$O$_3$ which is estimated to be ~1% of volume since no reflection is observed in the x-ray diffraction pattern. The Yb valence state was investigated on a bulk YbCo$_2$ sample by the X-ray absorption spectroscopy (XAS) for the Yb $L_3$-edge at the BL12XU (Taiwan Beamline), SPring-8, Japan. The partial fluorescence yield (PFY) mode was employed for the measurement, which allows a high-resolution evaluation. Details of the experimental setup were published elsewhere [23-25].

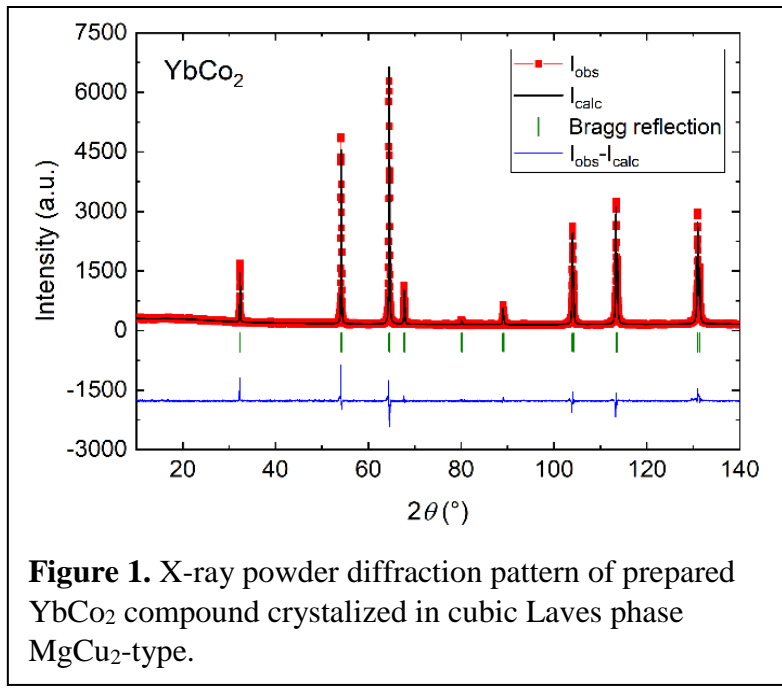

**Figure 1.** X-ray powder diffraction pattern of prepared YbCo$_2$ compound crystalized in cubic Laves phase MgCu$_2$-type.

The magnetization measurement was performed using the magnetic property measurement system with fields up to 7 T (MPMS, Quantum Design Inc.). Standard electrical resistivity and specific heat measurements were carried out by the Physical Property Measurement System (PPMS, Quantum Design Inc.) under magnetic field up to 7 T. Low-temperature measurement of specific heat down to 0.3 K was performed using $^3$He insert for PPMS.

## 3. Results

Figure 2 shows the PFY-XAS spectra of YbCo$_2$ measured at the Yb-$L_3$ absorption edge at $T$ = 300 K and 15.6 K. Three components are observed, which are contributions from the Yb$^{3+}$, Yb$^{2+}$, and the quadrupolar transition [23-25]. The results indicate that the component due to Yb$^{2+}$ is negligibly small. The Yb valence calculated from the spectra is 2.99+ and temperature-independent, suggesting an almost pure Yb$^{3+}$ state. This is in good agreement with the previously reported XAS result measured on the high-pressure synthesized YbCo$_2$ sample [21].

There is a peak shift in the incident photon energies of about 0.7 eV between the two measured temperatures. This shift was most prominent below 50 K, as is shown in the supporting information, Figure S1. Similar phenomena were observed in Yb compounds such as $YbInCu_4$ [24]. Although the mechanism is yet unclear, this possibly suggests a Fermi level shift with decreasing temperature.

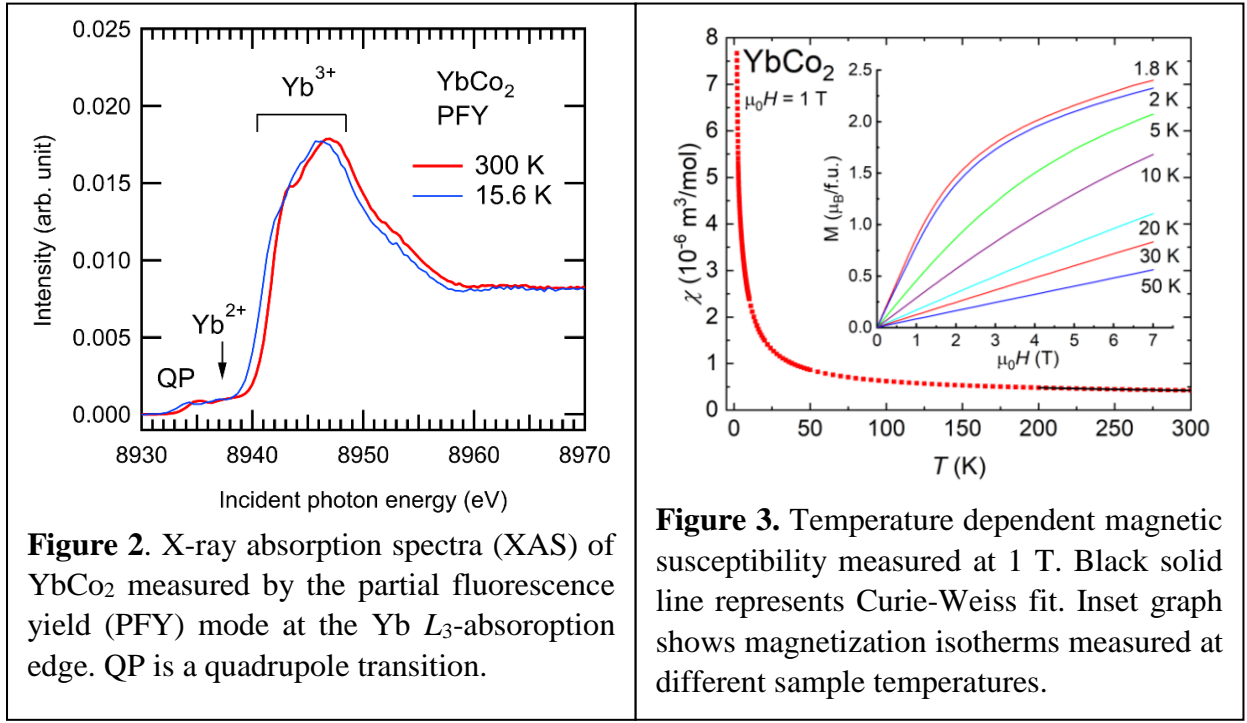

**Figure 2.** X-ray absorption spectra (XAS) of $YbCo_2$ measured by the partial fluorescence yield (PFY) mode at the Yb $L_3$-absoroption edge. QP is a quadrupole transition.

**Figure 3.** Temperature dependent magnetic susceptibility measured at 1 T. Black solid line represents Curie-Weiss fit. Inset graph shows magnetization isotherms measured at different sample temperatures.

Temperature dependence of the magnetic susceptibility measured at 1 T is depicted in Figure 3. There is no magnetic transition down to 2 K, although it was expected to occur at 5 K according to the previous theoretical prediction in Ref. [22]. The magnetic susceptibility has been analyzed with a modified Curie-Weiss law, $\chi(T) = C/(T - \theta_p) + \chi_0$, where $C$, $\theta_p$, and $\chi_0$ are the Curie constant, the Weiss temperature, and the temperature-independent susceptibility, respectively. The fitting in the temperature range from 200 K to 300 K gives an effective magnetic moment of $\mu_{eff} = 7.12(5)$ $\mu_B$ per formula unit. This value is much larger than that expected for a free $Yb^{3+}$ ion of 4.54 $\mu_B$. It indicates that Co $3d$-electrons also participate in the Curie-Weiss paramagnetism because the $Yb^{3+}$ valence state with magnetic moment was confirmed by X-ray absorption measurement. Thus, we assume the presence of $Yb^{3+}$ magnetic moment with the value of 4.54 $\mu_B$ and calculate the corresponding Co magnetic moment by the formula $\mu_{eff} = \sqrt{\mu_{Yb}^2 + 2\mu_{Co}^2}$ with the result $\mu_{Co} = 3.88$ $\mu_B$. This value is similar to the value of the Co-effective magnetic moment observed for $YCo_2$, $LuCo_2$, and $Y(Co_{1-x}Al_x)_2$ in Ref [17, 26]. The Curie-Weiss fit shows a negative Weiss temperature $\theta_p = -120$ K, pointing to an antiferromagnetic interaction in the paramagnetic region. On the other hand, in the case of itinerant electron magnetism, the negative $\theta_p$ is also seen for nearly ferromagnetic compounds such as $Y(Co_{1-x}Al_x)_2$ [26] and $Sr_{1-x}Ca_xRuO_3$ [27]. Thus, the contribution of Co $3d$ electrons may also be important in $\theta_p$ of $YbCo_2$ as well. In addition to that, possible Kondo interaction between conduction electrons and the $4f$ magnetic moments can yield negative $\theta_p$ in general. If the anti-parallel coupling between Co-3$d$ and Yb-4$f$ magnetic moments is favored, as is the case of HoCo$_2$, ErCo$_2$, etc., it can also give a negative $\theta_\text{p}$ on average. We should note that the crystalline electric field (CEF) also can affect the sign and values of $\theta_\text{p}$ and $\mu_\text{eff}$, although it was reported that the effect of CEF is not significant in the case of $R$Co$_2$ compounds [28]. From the specific heat measurements, as shown below, the CEF splitting is estimated to be at most 130 K. We thus employed the CW fitting temperature range of above 200 K so that the effect of CEF should be insignificant.

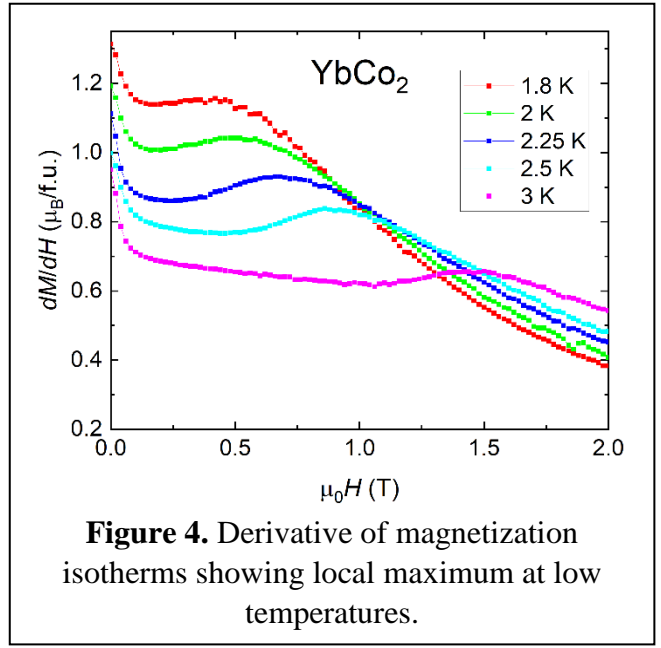

**Figure 4.** Derivative of magnetization isotherms showing local maximum at low temperatures.

The magnetic isotherms (see inset in Figure 3) demonstrate paramagnetic behavior for temperatures above 20 K. On the other hand, magnetization measurements at lower temperatures suggest nearly ferromagnetic-like behavior which can point to possible magnetic ordering at lower temperatures. The derivative of low-temperature part of magnetic isotherm shows a local maximum as depicted in Figure 4, which indicates the metamagnetic transition (MMT). The isostructural compounds YCo$_2$ and LuCo$_2$ show MMT out of paramagnetic state but the transition appears at a high magnetic field ~ 70 T and is caused by Co sublattice [29, 30]. This is quite contrasting with the present low-field metamagnetic behavior and the absence of step-like anomaly in the case of YbCo$_2$. On the other hand, it is notable that another classical example of a nearly-ferromagnetic itinerant-electron compound TiBe$_2$ with the same cubic Laves-type structure, exhibits a gradual

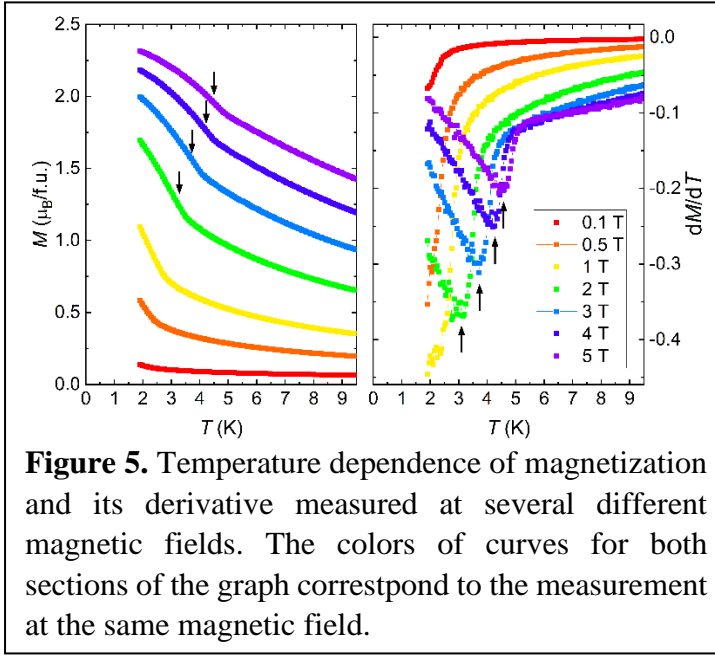

**Figure 5.** Temperature dependence of magnetization and its derivative measured at several different magnetic fields. The colors of curves for both sections of the graph correstpond to the measurement at the same magnetic field.

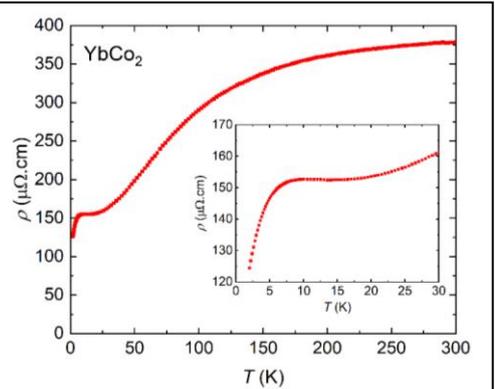

**Figure 6.** Temperature dependence of electrical resistivity. Inset graph shows resistivity in low temperature.

MMT in lower magnetic fields (~ 6 T), which is quite similar to the present observation [31,32].

Temperature dependence of magnetization demonstrates clear magnetic anomalies as depicted in Figure 5. Characteristic temperatures of the anomaly are taken as the minimum of the derivative. It is found that the application of a magnetic field shifts the anomaly to higher temperatures. These features resemble a field-induced ferromagnetic transition and are commonly observed regardless of whether the magnetism is based on itinerant-electrons (ErCo$_2$, MnAs, etc.) or local-moments (Gd$_5$Si$_2$Ge$_2$, PrInNi$_4$, etc.) [33-36]. On the other hand, it is notable that a similar evolution of field-induced phase transition has been reported for YbCo$_2$Zn$_{20}$ [2,37,38]. Its origin was interpreted by the level-crossing of crystal-field states caused by the application of the magnetic field. Similarities with YbCo$_2$ are discussed later.

In Figure 6, the temperature dependence of the electrical resistivity of YbCo$_2$ is depicted. The resistivity saturates at high temperatures. This is caused by the spin fluctuation scattering and spin disorder contribution to the electrical resistivity in the paramagnetic state as in other $R$Co$_2$ compounds [19]. With decreasing temperature, the resistivity gradually decreases below 200 K and reaches a local minimum at around 15 K. The decrease corresponds to the reduction of 4$f$-orbital degeneracy due to the CEF splitting. For further decreasing temperatures, the resistivity shows a small maximum at 8 K, below which the resistivity decreases rather rapidly. This rapid decrease can be attributed to another CEF splitting or the Kondo-lattice behavior, as is often observed in other Yb compounds [39]. Nevertheless, in comparison with other Yb-based compounds, the Kondo anomaly of YbCo$_2$ is situated at lower temperatures which indicates a very low Kondo temperature $T_K$ of a few kelvins.

Electrical resistivity measured under magnetic fields is shown in Figure 7. The value of the electrical resistivity is gradually suppressed by increasing the magnetic field and further, above 2 T, the change of curvature at low temperature suggests some kind of transition (see Figure 7 a). The derivative of electrical resistivity curves clearly shows a maximum, (Figure 7 b), which well corresponds to the magnetic anomalies shown in Fig. 5. The fit of the low-temperature evolution is depicted in Figure 7 a. The resistivity curve measured at 0 and 1 T has a concave character and can't be fitted by a power law $\rho = \rho_0 + AT^n$. Increasing the magnetic field induces power-law behavior in resistivity curves. The fit can be provided for curves measured above 1 T. As is depicted in Figure 7 a, the fit can be used for a wider temperature range with the increasing magnetic field. At the fields of 7 and 6 T, the electrical resistivity curves are fitted by the power law with the exponent $n \sim 1.6$. This seemingly indicates that

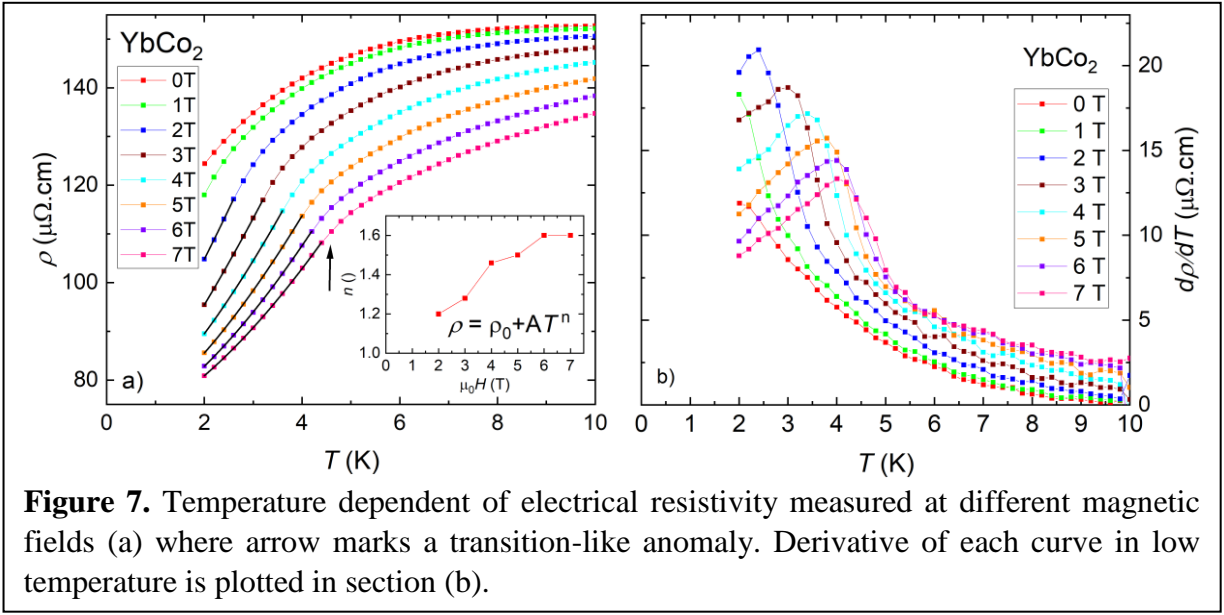

**Figure 7.** Temperature dependent of electrical resistivity measured at different magnetic fields (a) where arrow marks a transition-like anomaly. Derivative of each curve in low temperature is plotted in section (b).

YbCo$_2$ above 6 T is subjected to strong magnetic fluctuation which affects scattering processes of electrons, causing the breakdown of the normal Fermi-liquid behavior with $n = 2$. We note that the exponent $n = 5/3$ is predicted by spin fluctuation theory for itinerant ferromagnets [40-42]. With decreasing field, the deduced $n$ value is found to decrease as is shown in the inset of Figure 7 a. Although the precise evaluation becomes difficult for low fields because of the limited fitting range, the extrapolation of $n$ toward the zero field should likely result in further departing from the Fermi-liquid behavior ($n = 2$). This variation of $n$ by fields resembles the case of YbCu$_{5-x}$Au$_x$ and YbCu$_{5-x}$Al$_x$ [43,44], suggesting that YbCo$_2$ is possibly approaching the critical point in the zero field limit.

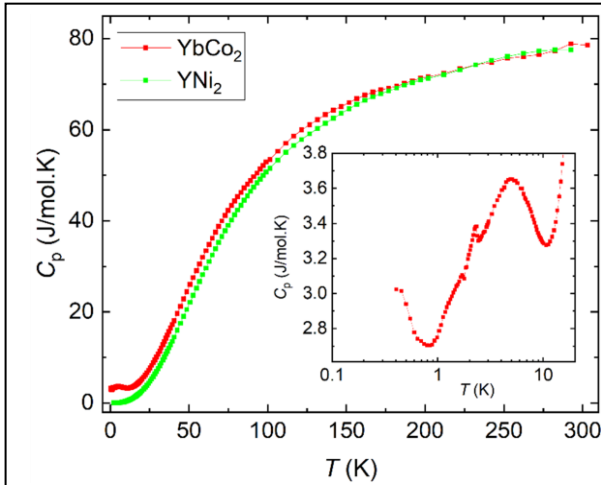

**Figure 8.** Specific heat measured on YbCo$_2$ (read squares) and non-magnetic analog YNi$_2$ (green squares). Inset shows detail of YbCo$_2$ specific heat in low temperature.

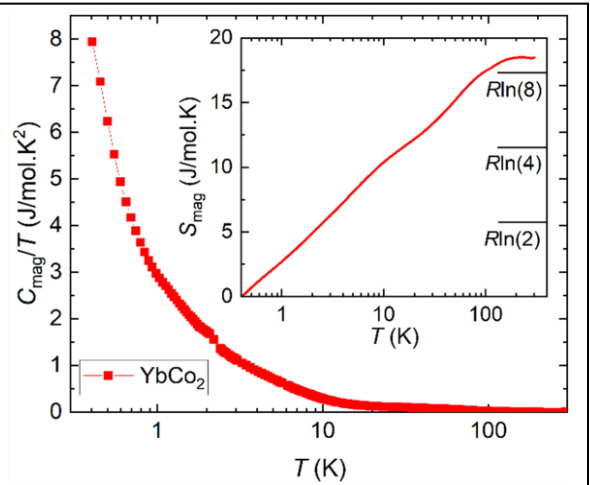

**Figure 9.** Temperature dependence (log scale) of magnetic specific heat $C_{mag}/T$ of YbCo$_2$. Inset shows calculated magnetic entropy $S_{mag}$ with marked several characteristic values.

The specific heat of YbCo$_2$ is depicted in Figure 8. A peak is observed around 2.3 K which originates from a small amount of Yb$_2$O$_3$. The data below 1 K show an increasing trend of specific heat, as is shown in the inset of Figure 8. The observed anomaly has a different temperature evolution than $T^{-2}$ expected for nuclear contribution, and calculated nuclear specific heat for each element in Ref [42] gives a much smaller value. Therefore, the increase in $C_p$ below 1 K is likely to be caused by electronic origin. The specific heat also shows a broad bump with a maximum at 4.9 K. This anomaly corresponds well to the broad maximum in the electrical resistivity.

The magnetic contribution to the specific heat, $C_{mag}$, has been evaluated by subtracting the specific heat of a nonmagnetic reference YNi$_2$. $C_p$ of YNi$_2$ below 2 K has been extrapolated by $C = \gamma T + \beta T^3$, where $\gamma$ and $\beta$ are constants (see Figure S2 in the supporting information). We attempted to compare the $C_{mag}$ with the CEF contribution and the Kondo model calculated by Rajan [45]. The results are shown in discussion section 4.1.

The $C_{mag}/T$ is depicted in Figure 9 where is observed the divergence character during cooling in logarithmic scale. Again, no signature of magnetic transition has been observed down to 0.3 K. Furthermore, $C_{mag}/T$ keeps increasing down to the lowest temperature measured, and the value exceeds 7 J/mol·K$^2$. This large value of $C_{mag}/T$ is comparable to those in YbPtBi [46], YbCo$_2$Zn$_{20}$ [47], YbCu$_{5-x}$Au$_x$ [48], and YbCu$_4$Ni [49], suggesting unusually massive electron state should be formed at lower temperatures in YbCo$_2$. The increasing character of $C_{mag}/T$ upon cooling agrees with the absence of Fermi liquid dependence ($\rho \sim T^2$) in the electrical resistivity, as described above.

These data have been used for a rough estimation of magnetic entropy, which is depicted in the inset of Figure 9. The magnetic entropy $S_{mag}$ reaches $R\ln(2)$ at ~ 2.5 K and shows a saturating tendency toward the value of $R\ln(4)$ at around 20 K. This temperature well corresponds to the temperature where $C_{mag}$ and $\rho$ show a local minimum. Finally, $S_{mag}$ increases to 150 K where saturates. The value of $R\ln(8)$ is reached at 100 K which points to the total entropy expected for $J = 7/2$ multiplet. Higher magnetic entropy than this value points to another contribution than from Yb, in our case Co magnetism.

Specific heat of YbCo$_2$ under the application of a magnetic field is depicted in Figure 10. The evolution of specific heat peaks in the magnetic field is clearly observed and it looks like a lambda-shaped anomaly for a second-order phase transition. Hence, these data strongly suggest the existence of a field-induced ordered phase, in agreement with what has been inferred by magnetization and electrical resistivity. The peak is moving to higher temperatures with increasing magnetic field, which possibly suggests the ferromagnetic character of this transition. If this is the case, YbCo$_2$ can be regarded as very close to ferromagnetic quantum criticality. Here, it is possible to estimate the electronic specific heat coefficient ($\gamma$) for the data in a magnetic field above 4 T by extrapolating $C_p/T$ data toward $T \rightarrow 0$ in Figure 10. The $\gamma$ values are roughly obtained to be 1 to 2 J/mol·K$^2$. Electrical resistivity indicates gradual restoration of Fermi liquid behavior by the magnetic field but it seems to be saturated at $n = 1.6$ pointing to ferromagnetic fluctuations.

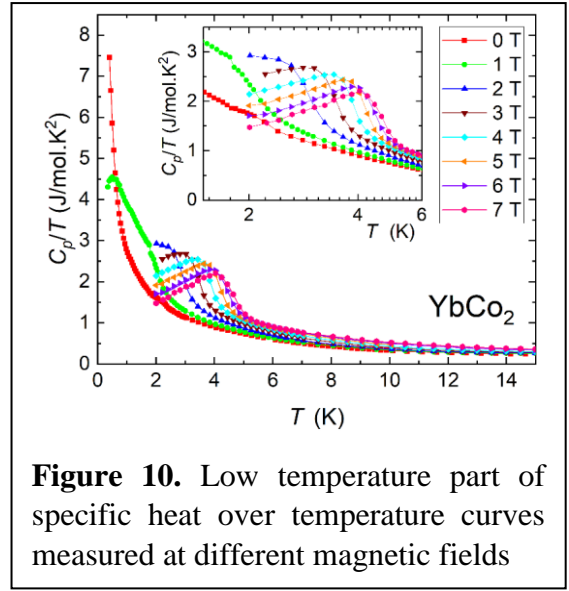

**Figure 10.** Low temperature part of specific heat over temperature curves measured at different magnetic fields

## 4. Discussion

*4.1. CEF splitting scheme*

Determination of the exact CEF splitting scheme for YbCo$_2$ is difficult due to the lack of single crystals. Here we attempt to make a rough evaluation of the CEF scheme using the specific heat results. According to the CEF calculation by K.R. Lea et al. in Ref. [50] ('The raising of angular momentum degeneracy of *f*-electron terms by cubic crystal fields'), $J = 7/2$ multiplet in cubic symmetry splits into two doublets, $\Gamma_6$ and $\Gamma_7$, and a quartet $\Gamma_8$. In particular, for x ~ 0.2, energies of the two doublets merge, whereby the $J = 7/2$ state splits into two quartets [50]. We then consider the possibility: (i) ground state quartet, with another quartet being the first excited state. Here, the quartet is either of $\Gamma_8$ or a quasi-quartet where two doublets are energetically very close to each other. (ii) Ground state doublet ($\Gamma_6$ or $\Gamma_7$), with the other doublet being the first excited state, and the quartet $\Gamma_8$ being the second excited state. (iii) Ground state doublet ($\Gamma_6$ or $\Gamma_7$), with quartet $\Gamma_8$ as the first excited state. Among them, case (iii) is excluded because this results in a too large $C_{mag}$ peak value of about 6 J/mol·K, which is not seen in the present case. Thus, we consider cases (i) and (ii).

In the case of (i), the maximum in $C_{mag}$ at $T = 50$ K is due to the CEF excitation from the ground state quartet to the excited state. The other maximum at $T = 5$ K could be attributed to the Kondo-singlet formation for the 4-fold degeneracy. The Kondo effect presumption lead us to employ the Bethe-Ansatz solution based on the Coqblin-Schrieffer (CS) model [45]. The results are shown in Figure 11 (a). The CEF splitting is estimated to be $\Delta \sim 130$ K. For the low-temperature Kondo peak at $T = 5$ K, 4-fold degenerated CS model almost explains the low-temperature data above 2 K. Notably, at temperatures below 2 K, the difference between the data and calculation is significant (hatched area). Since the degrees of freedom directly associated to the 4$f$ electrons are already taken into consideration in the fitting, the low-temperature enhancement in $C_{mag}$ can be due to other contributions such as the Co-3$d$ electron and its hybridization with 4$f$.

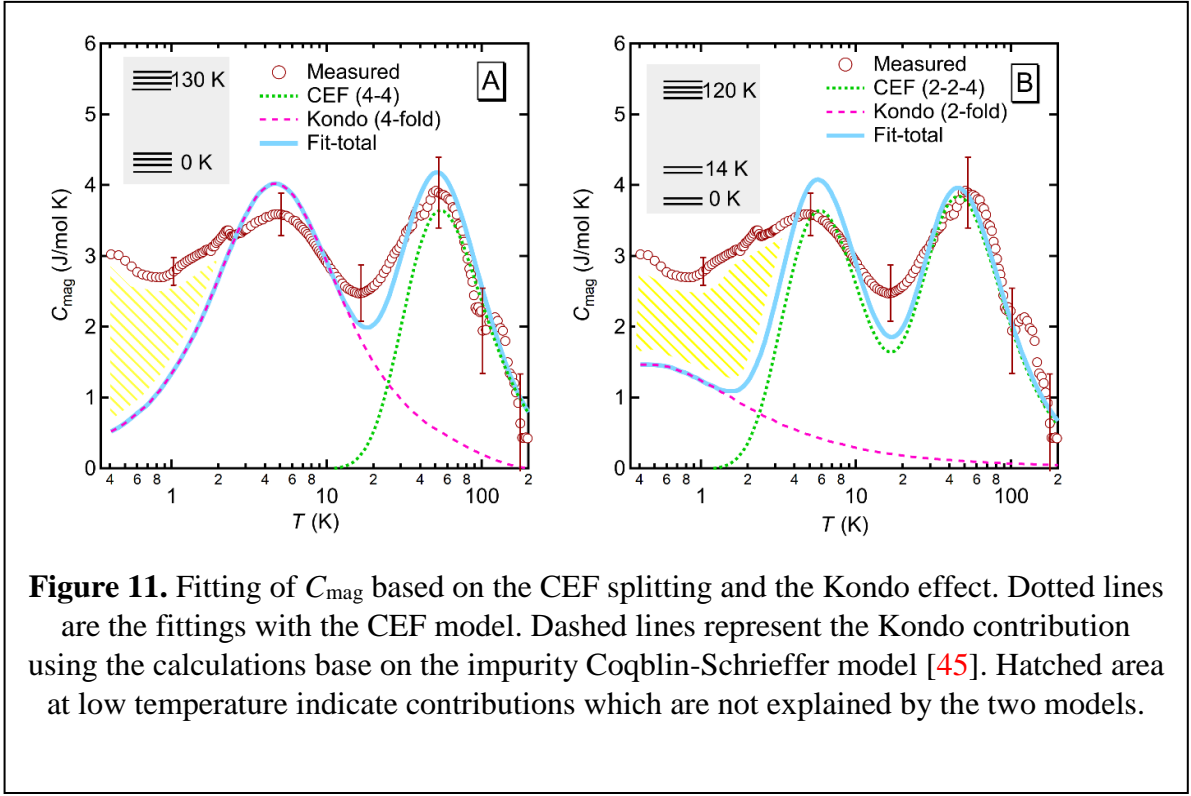

**Figure 11.** Fitting of $C_{mag}$ based on the CEF splitting and the Kondo effect. Dotted lines are the fittings with the CEF model. Dashed lines represent the Kondo contribution using the calculations base on the impurity Coqblin-Schrieffer model [45]. Hatched area at low temperature indicate contributions which are not explained by the two models.

Second, the splitting scheme (ii) is considered (see Figure 11 (b)). Here, the two peaks in $C_{mag}$ at $T = 50$ K and 5 K are attributed to the CEF excitations. The peak heights of about 4 J/mol·K are almost consistent with this splitting scheme. From the fitting, it is suggested that the second doublet is at 14 K and the quartet is at 120 K. Below $T = 5$ K, the data of $C_{mag}$ is significantly larger than the CEF calculation, thereby we attempted to fit the low-temperature part by using the CS model for a doublet ground state (2-fold degeneracy). In this case, the maximum value of $C_{mag}$ due to the CS model is only 1.4 J/mol·K, as depicted with the dashed line. Therefore, a large portion of the data below about 3 K (hatched area) is not explained solely by the 4$f$-electron contribution. Again, this part can be due to the magnetic correlation of the Co-3$d$ electron and its hybridization with 4$f$ electrons.

Thus, the low-temperature peak of $C_{mag}$ at 5 K is due either to the Kondo screening of a quartet or to the CEF splitting of two doublets. At present, we cannot determine which is the case. For both cases, it is suggested that the Kondo screening effect becomes prominent at low temperatures. As described before, the large $C_{mag}/T$ of YbCo$_2$ over 7 J/mol·K$^2$ is only comparable to a limited number of Yb compounds, such as YbPtBi, YbCo$_2$Zn$_{20}$, YbCu$_{5-x}$Au$_x$, and YbCu$_4$Ni [46-49]. Some of them were able to be roughly explained by a very low $T_K$ of the order of 1 K [2, 49]. However, in the present case of YbCo$_2$ the CEF and the Kondo model cannot explain the strong enhancement of $C_{mag}$ below about 2 K points to the new mechanisms of the electronic correlations related to 4$f$ and Co 3$d$ electrons.

*4.2. Magnetic-field-induced ordering*

The susceptibility measurements and the results of the Curie-Weiss fit gave a higher value of $\mu_{eff}$ than expected for Yb$^{3+}$. It leads us to infer the existence of the contribution of the Co magnetic moment. This is a contrasting case compared to the mixed valence state of CeCo$_2$, where the magnetic character of both Ce and Co is suppressed [10]. The metamagnetic-like behavior in magnetization brings a question about the similar origin of this behavior as has been observed in such heavy fermion metamagnetic compounds as CeRu$_2$Si$_2$, YbIr$_2$Zn$_{20}$, YbCu$_5$, UPd$_2$Al$_3$, URu$_2$Si$_2$, UTe$_2$, etc [1,2,39,51-55]. The common parameter in these compounds is the interplay of Kondo and RKKY interaction which results in the large electronic specific heat coefficient and change between local and itinerant magnetism by temperature or magnetic field. The observed low-temperature behavior of YbCo$_2$ differs from the above-mentioned compounds. The MMT in heavy fermion compounds usually moves to lower magnetic fields for higher temperatures, or it shows almost temperature-independent behavior. Contrary, our measurements of YbCo$_2$ manifest ferromagnetic-like properties point to a polarized paramagnetic phase or ferromagnetic transition hidden at a lower temperature than we could reach.

The data of specific heat witness a huge electronic specific heat coefficient in YbCo$_2$ roughly estimate in the range of 1 to 2 J/mol·K$^2$ for 4-7 T. This indicates that heavy fermion behavior evolves below the observed anomaly. This result suggests that the observed anomaly most likely involves not only the 3$d$ bands but the very heavy 4$f$ quasi-particle band as well. It can be supported by the results of the fit of electrical resistivity curves where non-Fermi liquid behavior is observed through the value of the exponent $n$. The critical state is also suggested by the Wilson ratio which is between 1.4-1.6 for measurements at 3-5 T. The value higher than 1 corresponds to the Fermi liquid state and less than 2 the Kondo state. It is to be noted that the observed values are in agreement with values of other heavy fermion compounds with strong correlations shown in Ref [56].

Our results from electrical resistivity, the temperature dependence of magnetization, and specific heat under magnetic field show pronounced anomalies. As discussed above, the behavior can be understood as field-induced transition, where both the metamagnetic behavior due to itinerant 3$d$ electrons and the heavy 4$f$ electrons can be involved. On the other hand, the anomalies are similar to reported observations on YbCo$_2$Zn$_{20}$ [2,37,38]. The field-induced ordered phase was reported in the measurement of the magnetic field along the <111> crystal lattice. This type of ordered phase is understood within the CEF model where the two Zeeman split levels occur a level-crossing at a finite magnetic field to form a quasi-doublet state, which causes an antiferro-quadrupole ordering [2]. This type of field-induced transition has similar features as those in the present observation of YbCo$_2$ (i) the increasing transition temperature with an increasing applied magnetic field, (ii) the ferromagnetic-like rapid increase of the magnetization without saturation below the ordered temperature, (iii) the same appearance of an anomaly in electrical resistivity and Fermi-liquid behavior below the anomaly under applied magnetic field. The specific heat analysis (section 4.1) brings another comparison by the possible appearance of two doublets presented which are separated with a small energy difference (~ 14 K). Since the exact CEF scheme is not yet determined for YbCo$_2$ at present, the occurrence of this field-induced level crossing and the resultant quadrupolar ordering is still an open question. Nevertheless, magnetic entropy change in the case of YbCo$_2$Zn$_{20}$ is ~ 0.6 J/mol·K [37] at 7 T whereas in YbCo$_2$ entropy change is ~ 6 J/mol·K at 5 T. The large changes in magnetic entropy seem to be in favor of magnetic ordering.

We summarized the magnetic-like phase diagram in Figure 12. The ferromagnetic-like behavior of the observed anomalies is very unusual in heavy fermion systems as most of the compounds undergo antiferromagnetic ordering. The heavy-fermion metamagnetism is likely coupled with the Co 3$d$-itinerant electron metamagnetism. The detailed study at low temperatures as well as neutron diffraction experiments under magnetic fields are necessary to resolve the origin of observed phenomena and investigation of grown single crystal would shed more light onto these issues as well.

## 5. Conclusions

The paper brings the first detailed study of YbCo$_2$ by the electrical resistivity, specific heat, and magnetization measurements. The Yb ion is indicated to be in the magnetic Yb$^{3+}$ valence state together with Co magnetic moment in the order of 3 $\mu_B$/Co. Low-temperature measurements revealed a broad bump ~ 8 K in measurements of electrical resistivity and specific heat which are ascribed to the Kondo contribution with low Kondo temperature. There was no indication of magnetic ordering down to 0.3 K, in contrast to the theoretically-predicted magnetic ordering at 5 K. $C_{mag}/T$ increases over 7 J/mol·K$^2$ with decreasing temperature, indicating that a very heavy fermion state evolves at lower temperatures. Moreover, this large part of the low-temperature $C_{mag}$ is not explained by the simple CEF and Kondo analysis, suggesting the unusual electronic correlation due to the interplay of 4$f$ and 3$d$ magnetism.

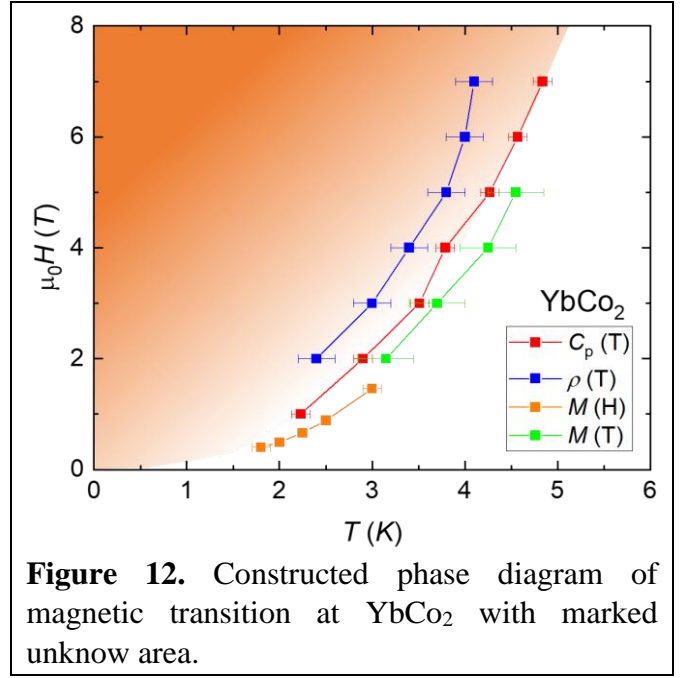

**Figure 12.** Constructed phase diagram of magnetic transition at YbCo$_2$ with marked unknow area.

The measurements under the application of a magnetic field revealed metamagnetic-like behavior and magnetic anomaly. The origin of the observed anomaly is discussed in three ways: (i) field-induced transition where the magnetic field polarized evidently strong magnetic correlations origins from 3$d$ and 4$f$ magnetic moments and; (ii) field-induced ordered phase caused by CEF level crossing. The representative compound with this ordered phase is YbCo$_2$Zn$_{20}$. In any case, the very large electronic specific heat coefficient below the field-induced observed anomaly suggests that the properties in YbCo$_2$ involve both: the itinerant electron magnetism of Co 3$d$-electrons and the Kondo effect within the vicinity of quantum criticality of Yb 4*f*-local moments. Partial support is brought by non-Fermi liquid behavior observed in electrical resistivity and from the Wilson ratio.

The results indicate that $YbCo_2$ is a unique and attractive example for the case where both the 3*d*- and 4*f*-electrons play a role in magnetic instability. Further experiments with single crystalline samples are desired.

**Acknowledgments**

This work was supported by the Grand-in-Aid from Japanese Society for the Promotion of Science, KAKENHI, Nos. 21F21322, 18K18743, and 22H01761. NT, HS, NT were supported by JST-MIRAI program, Grant No. JPMJMI18A3, Japan. TM acknowledges support from JST Mirai JPMJMI19A1. The specific heat measurement below 2 K was performed under the GIMRT Program of the Institute for Materials Research, Tohoku University (Proposal No. 202112-IRKAC-0029). The XAS measurements at BL12XU, SPring-8, were carried out under the proposal Nos. 2021B4253 and 2022A4258, corresponding NSRRC Proposals Nos.~2021-1-009 and 2021-2-104.